\documentclass{ws-procs9x6}
\begin{document}
%
%

\title{HIDDEN QUANTUM-MECANICAL SUPERSYMMETRY\\
IN EXTRA DIMENSIONS\footnote{
This review was given at 13th Regional Conference on Mathematical 
Physics, Antalya, Turkey, October 27-31, 2010.
}}
%
%
%
\author{Makoto Sakamoto}
%
%
%

\address{Department of Physics, Kobe University, \\
Rokkodai, Nada, Kobe 657-8501, Japan,\\
E-mail: dragon@kobe-u.ac.jp\\
}
%
%
%

%
%
\begin{abstract}
%
%
We study higher dimensional field theories with extra dimensions
from a 4d spectrum point of view.
It is shown that 4d mass spectra of spinor, gauge and gravity
field theories are governed by quantum-mechanical supersymmetry.
The 4d massless modes turn out to correspond to zero
energy vacuum states of the supersymmetry.
Allowed boundary conditions on extra dimensions compatible with
the supersymmetry are found to be severely restricted.

\end{abstract}
%
%
%

%
%
%

\bodymatter
%
%
%

%
%
%
\section{Introduction}
%
%
%
Gauge theories in higher dimensions are a promising candidate
beyond the Standard Model.
Such theories turn out to possess unexpectedly rich properties
that shed new light and give a deep understanding on high
energy physics. In fact, it has been shown that new mechanisms
of gauge symmetry breaking \cite{manton, fairlie, sherk,
hosotani, higgsless}, spontaneous supersymmetry 
breaking \cite{susy},
and breaking of translational invariance \cite{translation1,
translation2} can occur, and that various phase structures arise
in field theoretical models
on certain topological manifolds \cite{higgshosotani1, higgshosotani2,
highT}. 
Furthermore, new diverse scenarios of solving the hierarchy 
problem have been proposed\cite{rs, manybrane, 
nagasawasakamoto, sakamototakenaga1}.

Higher dimensional field theories will be described by 4d effective
theories at low energies.
Since we could not directly see extra dimensions, in particular, 
higher dimensional symmetries such as higher dimensional gauge
symmetry and general covariance symmetry,
one might ask what are remnants of the symmetries which originate 
from extra dimensions.
They have to be hidden in the 4d effective theories.
This is our motivation to investigate higher dimensional field
theories from a 4d mass spectrum point of view.
Our results show that the 4d mass spectrum is governed by
quantum-mechanical supersymmetry (QM SUSY).
Especially, the 4d massless spectrum is closely related to zero energy
vacuum states of the supersymmetry and depends crucially on boundary
conditions of extra dimensions, which are severely restricted
by compatibility with QM SUSY.

In higher dimensional scalar theories, QM SUSY would appear
in 4d spectrum but its appearance is found to be accidental.
In higher dimensional spinor, gauge and gravity theories,
QM SUSY always appears in 4d mass spectrum.
Its origin turns out to be chiral symmetry, higher dimensional
gauge symmetry and higher dimensional general covariance symmetry
for spinor, gauge and gravity theories, respectively.
It is interesting to note that all the symmetries guarantee the
masslessness of the fields.

The paper is organized as follows:
In section 2, we summarize the characteristic properties of QM SUSY.
In the subsequent sections, we examine the 4d mass spectrum of 
higher dimensional scalar, spinor, gauge and gravity theories,
separately and show that QM SUSY always appears in the 4d mass
spectrum except for scalar theories.
The  section 7 is devoted to conclusions.

%
%
%
\section{Minimal Supersymmetry Algebra}
%
%
%
In any higher dimensional spinor/gauge/gravity theories with extra
dimensions, quantum-mechanical supersymmetry turns out to be hidden
in 4d spectrum and to play an important role to determine
the spectrum of massless 4d fields, which are crucial ingredients
in constructing low energy effective theories.
The supersymmetric structure is found to be summarized in the 
minimal supersymmetry algebra, which consists of the
hermitian operators $H, Q$ and $F$, defined by
%
\begin{align}
\label{QM_SUSY_algebra_1}
H &= Q^{2},\\
\label{QM_SUSY_algebra_2}
(-1)^{F}\,Q &= -Q\,(-1)^{F},\\
\label{QM_SUSY_algebra_3}
(-1)^{F} &= 
  \left\{
   \begin{array}{ll}
     +1 & \quad{\textrm{for \lq\lq bosonic" states}},\\
     -1 & \quad{\textrm{for \lq\lq fermionic"  states}},
   \end{array} \right.
\end{align}
%
where $H$ and $Q$ are the Hamiltonian and the  supercharge.
The operator $F$ is called a \lq\lq fermion" number operator
and the eigenvalues of $(-1)^{F}$ are given by $+1$ for \lq\lq bosonic"
states and $-1$ for \lq\lq fermionic" ones, although
the words, boson and fermion, have nothing to do with
particles of integer spins and half-odd integer spins.
The readers should not confuse the quantum-mechanical supersymmetry
with supersymmetry in quantum field theory, which implies a symmetry
between bosonic states with integer spins and fermionic states
with half-odd integer spins.
The operators in the algebra 
(\ref{QM_SUSY_algebra_1})-(\ref{QM_SUSY_algebra_3}) are defined 
in quantum-mechanical
systems and hence the supercharge $Q$ does not possess any
spinor index.
We call the symmetry which obeys the algebra 
(\ref{QM_SUSY_algebra_1})-(\ref{QM_SUSY_algebra_3}) 
quantum-mechanical supersymmetry, or simply QM SUSY in this paper.

Let us first clarify characteristic properties of the  algebra
(\ref{QM_SUSY_algebra_1})-(\ref{QM_SUSY_algebra_3}),
We now show that if the system obeys the algebra, the spectrum
has the following properties:
%
\begin{itemize}
%
\item[1)]
The energy eigenvalues are non-negative, i.e. $E \ge 0$.
%
\item[2)]
Any positive energy state $\vert E,+\rangle$ of $(-1)^{F}=+1$
forms a pair with the state $\vert E,-\rangle$ of the same energy $E$ 
and $(-1)^{F}=-1$, and vice versa.
All positive energy states form supermultiplets.
%
\item[3)]
Zero energy states (if exist) do not necessarily form
pairs of supermultiplets.
\end{itemize}
%
Thus, a typical spectrum of QM SUSY systems will be given by Fig.1.
%
%
%
\begin{figure}
\begin{center}
\psfig{file=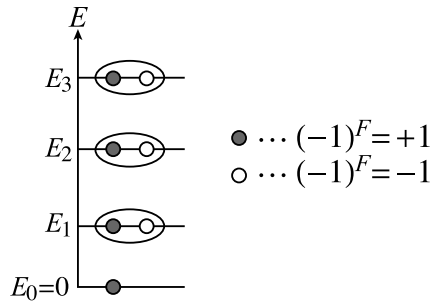,scale=0.35}
\end{center}
\vspace{-5mm}
\caption{A typical spectrum of QM SUSY}
\label{fig1}
\end{figure}
%
%
%

The first property 1) of $E\ge 0$ is derived from 
Eq.(\ref{QM_SUSY_algebra_1}) because
%
\begin{align}
\label{E>0}
E = \langle E\vert H \vert E \rangle
  = \langle E\vert Q^{2} \vert E \rangle
  = \vert\!\vert\,Q \vert E\, \rangle\vert\!\vert^{2}
  \ge 0
\end{align}
%
for any normalized energy eigenstate $\vert E\rangle$.
Here, we have used the facts that the supercharge $Q$ is hermitian
and the norm of any state is non-negative.
The second property 2) can be shown as follows:
Suppose that $\vert E,+\rangle$ is an energy eigenstate with
$(-1)^{F}=+1$.
Then, the state $Q\vert E,+\rangle$ has the same energy eigenvalue $E$
but the opposite eigenvalue of $(-1)^{F}$ because
$H (Q\vert E,+\rangle)
  = Q^{2} (Q\vert E,+\rangle)
  = Q H\vert E,+\rangle
  = E (Q\vert E,+\rangle)$
and
$(-1)^{F} (Q\vert E,+\rangle)
  = -Q (-1)^{F} \vert E,+\rangle
  = - (Q\vert E,+\rangle)$.
%
These relations imply that $Q\vert E,+\rangle \propto \vert E,-\rangle$.
Assuming $Q\vert E,\pm\rangle = \alpha_{\pm}\vert E,\mp\rangle$
with $\vert\!\vert\,\vert E,\pm\, \rangle\vert\!\vert^{2}=1$,
we find
%
\begin{align}
\label{normalization}
E = \langle E,\pm \vert H\vert E,\pm\, \rangle
  = \langle E,\pm \vert Q^{2}\vert E,\pm\, \rangle
  = \vert\!\vert\,Q\vert E,\pm\, \rangle\vert\!\vert^{2}
  = \vert \alpha_{\pm}\vert^{2}.
\end{align}
%
We can then take $\alpha_{\pm}=\sqrt{E}$ without any loss of 
generality.
Thus, the states $\vert E,+\, \rangle$ and $\vert E,-\, \rangle$
form a supermultiplet and are related each other through the
SUSY relations
%
\begin{align}
\label{SUSY_relations}
Q \vert E,\pm\, \rangle
 = \sqrt{E}\vert E,\mp\, \rangle.
\end{align}
%
The above result immediately shows that 
%
\begin{align}
\label{Q|E=0>=0}
Q\vert E,\pm\, \rangle = 0\quad \textrm{for}\ E=0.
\end{align}
%
This implies that zero energy states do not necessarily form
supermultiplets.

We have found that characteristic properties of QM SUSY is
nicely summarized in Fig.1.
Thus, if we encounter such a spectrum, QM SUSY is expected
to be hidden in the system.
In fact, we will find this type of spectrum again and again
in the following sections.

%
%
%
\section{5d Scalar}
%
%
%
Let us start with a 5d real massless scalar field theory compactified on
a circle $S^{1}$.
%
\begin{align}
\label{5d_scalar_action}
S = \int d^{4}x \int_{0}^{L} \hspace{-2mm}dy\,
    \Bigl\{ \frac{1}{2}\Phi(x,y)\bigl( 
      \partial^{\mu}\partial_{\mu} + \partial_{y}^{\ 2}\bigr)
      \Phi(x,y) - V(\Phi) \Bigr\},
\end{align}
%
where $x^{\mu}\ (\mu=0,1,2,3)$ denotes the 4-dimensional
Minkowski space-time coordinate and $y$ is the coordinate
of the extra dimension on the circle $S^{1}$ of the
circumference $L$.
The $V(\Phi)$ denotes a potential term but it will not
be concerned in our analysis.

Since the extra dimension is compactified on the circle of the
circumference $L$, we have to specify a boundary condition on
the field $\Phi(x,y)$.
Let us take a periodic boundary condition, as an example, i.e.
%
\begin{align}
\label{periodic_bc}
\Phi(x,y+L) = \Phi(x,y).
\end{align}
%
We will make a comment on other boundary conditions at the
end of this section.

In order to obtain the 4d mass spectrum, we expand the 5d
field $\Phi(x,y)$ into the Kaluza-Klein modes such as
%
\begin{align}
\label{mode_expansion_scalar}
\Phi(x,y) 
 = \phi_{0}^{(+)}(x)f_{0}^{(+)}(y)
   + \sum_{n=1}^{\infty} \bigl\{
     \phi_{n}^{(+)}(x)f_{n}^{(+)}(y)
     + \phi_{n}^{(-)}(x)f_{n}^{(-)}(y) \bigr\},
\end{align}
%
where $\phi_{n}^{(\pm)}(x)$ correspond to 4d scalar fields and
$f_{n}^{(\pm)}(y)$ are the mass eigenfunctions of the differential
operator $-\partial_{y}^{\ 2}$, i.e.
%
\begin{align}
\label{mode_function_1}
\left\{
   \begin{array}{l}
     f_{0}^{(+)}(y) = N_{0}^{(+)},\\
     f_{n}^{(+)}(y) = N_{n}^{(+)} \cos\bigl(\frac{2\pi y}{L}\bigr),\\
     f_{n}^{(-)}(y) = N_{n}^{(-)} \sin\bigl(\frac{2\pi y}{L}\bigr),
     \quad n=1,2,3\cdots.
   \end{array} \right.
\end{align}
%
Here, $N_{n}^{(\pm)}$ denote normalization constants.
We should note that the set of $\{f_{n}^{(\pm)}\}$ forms a complete
set so that the expansion (\ref{mode_expansion_scalar}) should be 
regarded as an identity.

Inserting the expansion (\ref{mode_expansion_scalar}) into the
action (\ref{5d_scalar_action}) and using the orthogonal relations
of $f_{n}^{(\pm)}(y)$ with an appropriate normalization, we find
%
\begin{align}
\label{4d_scalar_action}
S = &\int d^{4}x
    \Bigl\{ 
      \frac{1}{2}\phi_{0}^{(+)}(x)\partial^{\mu}\partial_{\mu}
                 \phi_{0}^{(+)}(x)
    + \frac{1}{2}\sum_{n=1}^{\infty} \Bigl(      
      \phi_{n}^{(+)}(x)
                 \bigl(\partial^{\mu}\partial_{\mu}-m_{n}^{\ 2}\bigr)
                 \phi_{n}^{(+)}(x)\notag\\
    &\qquad\ + \frac{1}{2}\phi_{n}^{(-)}(x)
                 \bigl(\partial^{\mu}\partial_{\mu}-m_{n}^{\ 2}\bigr)
                 \phi_{n}^{(-)}(x) \Bigr)  - V(\phi)                
    \Bigr\},
\end{align}
%
where $m_{n}$ is the 4d mass of the field $\phi_{n}^{(\pm)}$
and is given by
 %
\begin{align}
\label{KK_mass}
m_{n} = \frac{2\pi n}{L},\quad n=0,1,2,\cdots.
\end{align}
%
It follows that there appears a single massless mode $\phi_{0}^{(+)}$
and that all massive modes $\phi_{n}^{(\pm)}\ (n=1,2,\cdots)$
are doubly degenerate.
Thus, the 4d mass spectrum is given by Fig.2.
%
%
%
\begin{figure}
\begin{center}
\psfig{file=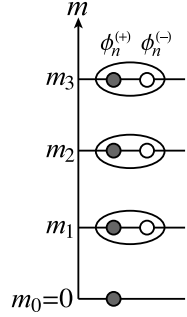,scale=0.35}
\end{center}
\vspace{-5mm}
\caption{Spectrum of 5d scalar on $M^{4}\times S^{1}$}
\label{fig2}
\end{figure}
%
%
%
%
This is nothing but a typical QM SUSY spectrum!
We can, in fact, show that the minimal supersymmetry algebra
appears in the system and the model will be the simplest
higher dimensional field model that possesses QM SUSY.

Now the question is what are the operators
$H, Q$ and $(-1)^{F}$ of QM SUSY in the present system.
The answer is
 %
\begin{align}
\label{QM_SUSY_scalar}
H=-\partial_{y}^{\ 2},\ \ \ Q=-i\partial_{y},\ \ \ (-1)^{F}={\cal P}.
\end{align}
%
The eigenvalues of the Hamiltonian $H$ correspond to the mass
squared $m_{n}^{\ 2}$.
The supercharge Q is just the momentum operator, i.e. $Q=-i\partial_{y}$
and satisfies the desired relation $H=Q^{2}$.
The operator $(-1)^{F}$ is given by the parity operator ${\cal P}$.
It implies that the states of $(-1)^{F}=+1\ (-1)$ correspond
to even (odd) parity states.
It is easy to verify that $Q$ anticommutes with $(-1)^{F}$, 
as they should.
The parity even (odd) function $f_{n}^{(+)}(y)\ (f_{n}^{(-)}(y))$
has $(-1)^{F}=+1\ (-1)$ and they form a supermultiplet because
$Qf_{n}^{(\pm)}(y)=-i\partial_{y}f_{n}^{(\pm)}(y)\propto f_{n}^{(\mp)}(y)$ 
for $n>0$.
Furthermore, $Qf_{0}^{(+)}=0$ because $f_{0}^{(+)}$ is independent
of $y$.
This implies that the zero mode $f_{0}^{(+)}$ has no superpartner, 
as expected.
Therefore, we have confirmed that the degeneracy for the nonzero
modes in the  4d mass spectrum (see Fig.2) can be  explained by QM SUSY.

As mentioned before, we make a comment on boundary conditions.
In the above analysis, we assumed the periodic boundary condition
(\ref{periodic_bc}).
We can show that the system with the antiperiodic boundary condition
$\Phi(y+L)=-\Phi(y)$ possesses QM SUSY as well but without
any massless 4d state.
Boundary conditions other than periodic and antiperiodic boundary
conditions,\footnote{
Examples of other boundary conditions are Dirichlet boundary
condition (b.c.) $\Phi(0)=0$, Neumann b.c. $\partial_{y}\Phi(0)=0$,
twisted b.c. $\Phi(y+L) = e^{i\theta}\Phi(y)$ for a complex scalar.
}
however, lead to non-degenerate 4d spectrum and 
no QM SUSY.\footnote{
The systems with periodic or antiperiodic boundary condition
can have an accidental symmetry, i.e. parity symmetry.
This is the origin of QM SUSY as well as the degeneracy in the 4d spectrum.
}
Thus, we conclude that QM SUSY found in the scalar field theory is 
accidental and there is no general mechanism to guarantee QM SUSY
in any scalar field theories.

%
%
%
\section{5d Spinor}
%
%
%
In this section, we consider a 5d spinor filed on an interval
$(0\le y\le L)$:
%
\begin{align}
\label{5d_spinor_action}
S = \int d^{4}x \int_{0}^{L}\hspace{-2mm}dy\,
     \bar{\Psi}(x,y) \Bigl(
      i\gamma^{\mu}\partial_{\mu} + i\gamma^{y}\partial_{y}
      + M + \lambda \varphi(y) \Bigr) \Psi(x,y),
\end{align}
%
where $\Psi(x,y)$ is a 4-component 5d Dirac spinor field
and $M$ is a 5-dimensional (bulk) mass.
The $\gamma^{y}$ is given by $\gamma^{y}\equiv \gamma^{0}
\gamma^{1}\gamma^{2}\gamma^{3} = -i\gamma^{5}$.
Here, we have introduced a coupling to a real scalar field $\varphi(y)$
and allow it to have a nontrivial $y$-dependence
as a background field.
The spinor $\Psi(x,y)$ can be expanded as 
%
\begin{align}
\label{mode_expansion_spinor}
\Psi(x,y) 
 = \sum_{n} \bigl\{
     \psi_{+,n}(x)f_{n}(y) + \psi_{-,n}(x)g_{n}(y) \bigr\},
\end{align}
%
where $\psi_{\pm,n}$ are 4-dimensional chiral spinors defined by
$\gamma^{5}\psi_{\pm,n}=\pm\psi_{\pm,n}$.
The sets of functions $\{f_{n}(y)\}$ and $\{g_{n}(y)\}$
are assumed separately to form complete sets and should be chosen for
$\psi_{\pm,n}(x)$ to be 4d mass eigenstates.
From the representation theory of the Poincar{\' e} group,
a massive 4d Dirac spinor $\psi_{n}$ consists of chiral
spinors $\psi_{+,n}$ and $\psi_{-,n}$ and they form the mass terms
$m_{n}\bar{\psi}_{\pm,n}\psi_{\mp,n}$.
On the other hand, a massless 4d spinor is chiral and hence
does not necessarily form a pair of $\psi_{+,0}$ and $\psi_{-,0}$.
Therefore, the 4d mass spectrum of (infinitely many) 4d spinors
$\{\psi_{\pm,n}\}$ will be schematically given by Fig.3.
%
%
%
\begin{figure}
\begin{center}
\psfig{file=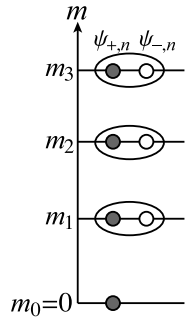,scale=0.35}
\end{center}
\vspace{-5mm}
\caption{Spectrum of 5d spinor}
\label{fig3}
\end{figure}
%
%
%
%
%
%
%
%
%
%
%
This is nothing but a typical QM SUSY spectrum, as discussed in
section 2.
Thus, we expect that the minimal supersymmetry algebra is hidden
in the 4d spectrum.
This is indeed the case, as we will see below.

The 5d Dirac equation for $\Psi(x,y)$ is given by
%
\begin{align}
\label{5d_Dirac_eq}
\bigl[ i\gamma^{\mu}\partial_{\mu} + \gamma^{5}\partial_{y}
       + M + \lambda\varphi(y) \bigr]\Psi(x,y) = 0. 
\end{align}
%
In terms of $\psi_{\pm,n}$, the above equation can be 
decomposed as 
\begin{align}
\label{5d_Dirac_eq_+}
&\sum_{n}\bigl( i\gamma^{\mu}\partial_{\mu}\psi_{+,n}(x) \bigr) f_{n}(y)
 + \sum_{n}\psi_{-,n}(x) \bigl({\cal D}^{\dagger}g_{n}(y)\bigr)
 = 0,\\
\label{5d_Dirac_eq_-}
&\sum_{n}\bigl( i\gamma^{\mu}\partial_{\mu}\psi_{-,n}(x) \bigr) g_{n}(y)
 + \sum_{n}\psi_{+,n}(x) \bigl({\cal D}f_{n}(y)\bigr)
 = 0,
\end{align}
%
where
%
\begin{align}
\label{def_D&D+}
{\cal D}
 = \partial_{y} + M +\lambda\varphi(y),\quad
{\cal D}^{\dagger}
 &= -\partial_{y} + M +\lambda\varphi(y).
\end{align}
%
We then require $f_{n}(y)$ an $g_{n}(y)$ to be the eigenfunctions
of the differential operators ${\cal D}^{\dagger}{\cal D}$ and
${\cal D}{\cal D}^{\dagger}$, respectively, i.e.
%
\begin{align}
\label{D+Df=m^2f}
{\cal D}^{\dagger}{\cal D} f_{n}(y)
 &= m_{n}^{\ 2} f_{n}(y),\\
\label{DD+g=m^2g}
{\cal D}{\cal D}^{\dagger} g_{n}(y)
 &= m_{n}^{\ 2} g_{n}(y).
\end{align}
%
Since ${\cal D}^{\dagger}{\cal D}$ and
${\cal D}{\cal D}^{\dagger}$ are hermitian,\footnote{
Boundary conditions for $f_{n}(y)$ and $g_{n}(y)$
have to be chosen for ${\cal D}^{\dagger}$ to be 
hermitian conjugate to ${\cal D}$.
We will discuss how to determine boundary conditions later.
}
the sets of $\{f_{n}(y)\}$ and $\{g_{n}(y)\}$ form
complete sets, as they should.
It follows from Eqs.(\ref{D+Df=m^2f}), (\ref{DD+g=m^2g}) that 
${\cal D}f_{n}$ (${\cal D}^{\dagger}g_{n}$) obeys the
same eigenequation (\ref{DD+g=m^2g}) ((\ref{D+Df=m^2f})) as
$g_{n}$ ($f_{n}$), and hence that $f_{n}$ and $g_{n}$ are 
related each other through the SUSY relations
%
\begin{align}
\label{spinor_SUSY_relation_1}
m_{n}g_{n}(y)
 &= {\cal D}f_{n}(y),\\
\label{spinor_SUSY_relation_2}
m_{n}f_{n}(y)
 &= {\cal D}^{\dagger} g_{n}(y),
\end{align}
%
with appropriate normalizations.
Thus the eigenvalues of $f_{n}$ and $g_{n}$ are doubly degenerate
(except for $m_{n}=0$), as expected.

The minimal supersymmetry algebra is manifest by introducing
the operators as
%
\begin{align}
\label{def_H_spinor}
H = Q^{2} 
   = \left(
      \begin{array}{cc}
       {\cal D}^{\dagger}{\cal D} & 0 \\
       0 & {\cal D}{\cal D}^{\dagger}
      \end{array} \right) ,\quad
Q = \left(
      \begin{array}{cc}
       0 & {\cal D}^{\dagger}\\
       {\cal D} & 0
      \end{array} \right) ,\quad
(-1)^{F}
  = \left(
      \begin{array}{cc}
       1 & 0\\
       0 & -1
      \end{array} \right) .
\end{align}
%
Those operators  act on 2-component wavefunctions
%
\begin{align}
\label{2_component_wavefunction}
\vert \Psi \rangle 
 = \left(
    \begin{array}{c}
     f(y)\\
     g(y)
    \end{array} \right).
\end{align}
%

Now, it is not difficult to show that with the SUSY relations
(\ref{spinor_SUSY_relation_1}), (\ref{spinor_SUSY_relation_2}) and with 
the orthonormal relations of $\{f_{n}(y)\}$ and $\{g_{n}(y)\}$
the action can be written, in terms of the 4d spinors, into the form
%
\begin{align}
\label{spinor_action_2}
S = \int d^{4}x \bigl\{ {\cal L}_{m=0} + {\cal L}_{m\ne 0}\bigr\},
\end{align}
%
where ${\cal L}_{m=0}$ is the part of the Lagrangian consisting
of massless chiral spinors and 
%
\begin{align}
\label{L_{m =/ 0}}
{\cal L}_{m\ne 0}
 = \sum_{m_{n}\ne 0} \bar{\psi}_{n}(x) (i\gamma^{\mu}\partial_{\mu}
     + m_{n}) \psi_{n}(x)
\end{align}
%
with $\psi_{n} = \psi_{+,n}+\psi_{-,n}$ for $m_{n}\ne 0$.
Thus, we have shown that $\psi_{\pm,n}$ are mass eigenstates
with $m_{n}$, as announced before.

To determine the chiral zero mode part ${\cal L}_{m=0}$,
we need to specify boundary conditions at $y=0,L$ for 
$f_{n}(y)$ and $g_{n}(y)$.
It turns out that the choice of boundary conditions is 
crucial for the existence of massless chiral spinors.
Allowed boundary conditions compatible with QM SUSY have been
classified and are listed below:\cite{bc_QMSUSY_1, bc_QMSUSY_2}
%
%
\begin{itemize}
%
\item[i)] 
${\cal D}f_{n}(0)=0={\cal D}f_{n}(L),\ g_{n}(0)=0=g_{n}(L)$,
%
\item[ii)] 
$f_{n}(0)=0=f_{n}(L),\ {\cal D}^{\dagger}g_{n}(0)=0
={\cal D}^{\dagger}g_{n}(L)$,
%
\item[iii)] 
${\cal D}f_{n}(0)=0=f_{n}(L),\ g_{n}(0)=0={\cal D}^{\dagger}g_{n}(L)$,
%
\item[iv)] 
$f_{n}(0)=0={\cal D}f_{n}(L),\ {\cal D}^{\dagger}g_{n}(0)=0=g_{n}(L)$.
\end{itemize}
%
%
Since the mode functions $f_{n}(y)$ and $g_{n}(y)$ obey the SUSY
relations (\ref{spinor_SUSY_relation_1}) and 
(\ref{spinor_SUSY_relation_2}),
chiral zero modes (if any) should satisfy
%
\begin{align}
\label{f_0}
{\cal D}f_{0}(y) &= 0,\\
\label{g_0}
{\cal D}^{\dagger}g_{0}(y) &= 0,
\end{align}
%
with $m_{0}=0$.
These first order differential equations can easily be solved as
%
\begin{align}
\label{solution_f_0}
f_{0}(y) 
 &= N_{0} \exp \Bigl\{
     -\int_{0}^{y}dy' \bigl( M + \lambda\varphi(y')\bigr)\Bigr\},\\
\label{solution_g_0}
g_{0}(y) 
 &= \bar{N}_{0} \exp \Bigl\{
     +\int_{0}^{y}dy' \bigl( M + \lambda\varphi(y')\bigr)\Bigr\}.
\end{align}
%
We should emphasize that the above solutions do not insure the 
existence of the massless chiral spinors $\psi_{+,0}$ and $\psi_{-,0}$
because they have to be discarded from the physical spectrum 
if $f_{0}(y)$ and/or $g_{0}(y)$ do not obey the boundary conditions.
It is easy to see that $f_{0}(y)$ given in (\ref{solution_f_0})
obeys the boundary conditions only for i) and that $g_{0}(y)$ 
in (\ref{solution_g_0}) obeys them only for ii).
Therefore, we find that
%
\begin{align}
\label{L_{m=0}}
{\cal L}_{m=0}
 = \left\{
    \begin{array}{ll}
     \bar{\psi}_{+,0}(x)i\gamma^{\mu}\partial_{\mu}\psi_{+,0}(x) 
     & \quad{\textrm{for i)}},\\
     \bar{\psi}_{-,0}(x)i\gamma^{\mu}\partial_{\mu}\psi_{-,0}(x) 
     & \quad{\textrm{for ii)}},\\
     0
     & \quad{\textrm{for iii) and iv)}}.\
    \end{array} \right.
\end{align}
%

The extension of the above analysis to higher dimensions $M^{4}\times K^{N}$
will be straightforward.
The $\Gamma$-matrices on $M^{4}\times K^{N}$ may be constructed,
in terms of the $\gamma$-matrices on $M^{4}$ and the $\bar{\gamma}$-matrices
on $K^{N}$, as 
%
\begin{align}
\label{def_Gamma_natrices}
\Gamma^{\mu} 
 &= \gamma^{\mu} \otimes I_{2^{[N/2]}}, & \mu=0,1,2,3,\notag\\
\Gamma^{i}
 &= \gamma^{5} \otimes \bar{\gamma}^{i}, & i=1,2,\cdots,N,
\end{align}
%
which satisfy
%
\begin{align}
\label{Cliford_algebra_Gamma}
\{\Gamma^{\mu}, \Gamma^{\nu}\} 
 &= -2 \eta^{\mu\nu} I_{4}\otimes I_{2^{[N/2]}}, & \mu,\nu=0,1,2,3,\notag\\
\{\Gamma^{i}, \Gamma^{j}\} 
 &= -2 \delta^{ij} I_{4}\otimes I_{2^{[N/2]}}, & i,j=1,2,\cdots,N,\notag\\
\{\Gamma^{\mu}, \Gamma^{j}\} 
 &= 0.
\end{align}
%
Here, $[N/2]$ denotes the Gauss symbol and $I_{n}$ is the
$n\times n$ identity matrix.
The structure of the $\Gamma$-matrices may imply that 
a $(4+N)$-dimensional spinor $\Psi(x,y)$ can be expanded as
%
\begin{align}
\label{(4+N)d_spinor}
\Psi(x,y)
 = \sum_{n}\bigl\{
   \psi_{+,n}(x)\otimes \xi_{+,n}(y)
   + \psi_{-,n}(x)\otimes \xi_{-,n}(y) \bigr\},
\end{align}
%
where $\psi_{\pm,n}(x)$ ($\xi_{\pm,n}(y)$) denote 
4-dimensional ($N$-dimensional) spinors and $x^{\mu}$ ($y^{i}$) are
the coordinates of $M^{4}$ ($K^{N}$).
The 4d mass spectrum of $\psi_{\pm,n}$ will be schematically
given just like Fig.3 and the mass eigenfunctions
$\xi_{+,n}(y)$ and $\xi_{-,n}(y)$ will form a supermultiplet, 
though we will not proceed further.

%
%
%
\section{5d Vector}
%
%
%
In this section, we consider a (4+1)-dimensional abelian gauge theory 
on an interval $(0\le y \le L)$:
%
\begin{eqnarray}
S=\int d^4 x \int_0^L dy \sqrt{-g(y)} \left\{ -\frac{1}{4} 
F_{MN}(x,y)F^{MN}(x,y) \right\}
\end{eqnarray}
%
with a non-factorizable metric
%
\begin{equation}
ds^2 =e^{-4W(y)}\eta_{\mu \nu}dx^{\mu}dx^{\nu} +g_{55}(y)dy^2.
\end{equation}
%
The metric reduces to the warped metric discussed by Randall and 
Sundrum\cite{rs} when $g_{55}(y)=1$ and $W(y)=\frac{1}{2}k|y|$.
Another choice of $g_{55}(y)=e^{-4W(y)}$ leads to the model 
discussed in Ref.[\refcite{nagasawasakamoto}], in which a 
hierarchical mass spectrum has been observed.

In order to expand the 5d gauge fields $A_{\mu}(x,y)$ and $A_{y}(x,y)$ 
into 4d mass eigenstates and to make a QM SUSY structure manifest,
we introduce the operators $H, Q$ and $(-1)^{F}$ 
as follows:\cite{SUSY_gauge_1, SUSY_gauge_2}
%
\begin{eqnarray}
&&H=Q^2 =\left(
	\begin{array}{cc}
		-\frac{1}{\sqrt{g_{55}}}\partial_y \frac{e^{-4W}}{\sqrt{g_{55}}}
		\partial_y &0 \\
		0& -\partial_y\frac{1}{\sqrt{g_{55}}}\partial_y 
		\frac{e^{-4W}}{\sqrt{g_{55}}}
	\end{array}
	\right), \\
&&Q=\left(
		\begin{array}{cc}
			0 & -\frac{1}{\sqrt{g_{55}}}\partial_y \frac{e^{-4W}}{\sqrt{g_{55}}} \\
			\partial_y & 0
		\end{array}
	\right),\\
&&(-1)^{F}=\left(
		\begin{array}{cc}
			1 & 0 \\
			0 & -1
		\end{array}
	\right),
\end{eqnarray}
%
which act on two-component vectors
%
\begin{equation}
	|\Psi \rangle =\left(
		\begin{array}{c}	
			f(y) \\
			g(y)
		\end{array}
	\right).
\end{equation}
The inner product of two states $|\Psi_1\rangle$ and $|\Psi_2 \rangle$ 
is defined by 
%
\begin{eqnarray}
\label{vector_inner_product}
	\langle \Psi_2 | \Psi_1 \rangle =\int_0^L dy \sqrt{g_{55}(y)} \left\{
		f_2(y) f_1(y) + \frac{e^{-4W(y)}}{g_{55}(y)}g_2(y)g_1(y) \right\}.
\end{eqnarray}
%
To obtain consistent boundary conditions for the functions $f(y)$ 
and $g(y)$ in $|\Psi\rangle$, we first require that the supercharge 
$Q$ is hermitian with respect to the inner product 
(\ref{vector_inner_product}), i.e.
%
\begin{equation}
	\langle \Psi_2 | Q\Psi_1 \rangle =\langle Q \Psi_2 | \Psi_1 \rangle.
\end{equation}
%
It turns out that the functions $f(y)$ and $g(y)$ have to obey one 
of the following four types of boundary conditions:
%
\begin{eqnarray}
	& {\rm i)} & g(0) =g(L) =0 , \\
	& {\rm ii)} & f(0)=f(L)=0,\\
	& {\rm iii)} & g(0)=f(L)=0, \\
	& {\rm iv)} & f(0)=g(L)=0. 
\end{eqnarray}
%
We further require that the state $Q |\Psi\rangle$ obeys the same 
boundary conditions as $|\Psi\rangle$, otherwise $Q$ is not a well 
defined operator and \lq\lq bosonic" and \lq\lq fermionic" states 
would not form supermultiplets.
The requirement leads to 
%
\begin{align}
	& \partial_y f(0) =\partial_y f(L) =0  &{\rm for}\ {\rm i)},\\
	& \partial_y \left( \frac{e^{-4W}}{\sqrt{g_{55}}} g \right)(0) =\partial_y \left( \frac{e^{-4W}}{\sqrt{g_{55}}} g \right)(L) =0 &{\rm for} \ {\rm ii)}, \\
	&\partial_y f(0)=\partial_y \left( \frac{e^{-4W}}{\sqrt{g_{55}}} g \right)(L)=0 
	&{\rm for}\ {\rm iii)}, \\
	&\partial_y  \left( \frac{e^{-4W}}{\sqrt{g_{55}}} g \right)(0) 
	=\partial_y f(L)=0 &{\rm for}  \ {\rm iv)}.
\end{align}
%
Combining all the above results, we have found the four types 
of boundary conditions compatible with supersymmetry,\cite{SUSY_gauge_1, SUSY_gauge_2}
%
\begin{eqnarray}
\label{bc_gauge_1}
	&&{\rm Type}\ ({\rm N,N}) : \left\{	
		\begin{array}{l}
			\partial_y f(0) =\partial_y f(L) =0 , \\
			g(0) =g(L) =0,
		\end{array}
	\right. \\
	&&{\rm Type}\ ({\rm D,D}) : \left\{	
		\begin{array}{l}
			f(0)=f(L)=0, \\
\label{bc_gauge_2}
			\partial_y \left( \frac{e^{-4W}}{\sqrt{g_{55}}} g \right)(0)=\partial_y \left( \frac{e^{-4W}}{\sqrt{g_{55}}} g \right)(L)=0,
			
		\end{array}
	\right. \\
\label{bc_gauge_3}
	&&{\rm Type}\ ({\rm N,D}) : \left\{	
		\begin{array}{l}
			\partial_y f(0) = f(L) =0 , \\
			g(0) =\partial_y \left( \frac{e^{-4W}}{\sqrt{g_{55}}} g \right)(L)=0,
		\end{array}
	\right. \\
	&&{\rm Type}\ ({\rm D,N}) : \left\{	
		\begin{array}{l}
			f(0) = \partial_y f(L) =0 , \\
\label{bc_gauge_4}
			\partial_y \left( \frac{e^{-4W}}{\sqrt{g_{55}}} g \right)(0)=g(L)=0.
		\end{array}
	\right. 
\end{eqnarray}
%
It follows that the above boundary conditions ensure the hermiticity 
of the Hamiltonian, i.e.
%
\begin{equation}
	\langle \Psi_2 |H \Psi_1 \rangle =\langle H \Psi_2 | \Psi_1 \rangle.
\end{equation}
%
Therefore, we have succeeded to obtain the consistent set of boundary 
conditions that ensure the hermiticity of the supercharge and the 
Hamiltonian and also that the action of the supercharge on 
$|\Psi\rangle$ is well defined.
Since the supersymmetry is a direct consequence of higher-dimensional 
gauge invariance, our requirements on boundary conditions should be, 
at least, necessary conditions to preserve it.
It turns out that the boundary conditions obtained above are consistent 
with those in Ref.[\refcite{higgsless}], although it is less obvious how the 
requirement of the least action principle proposed 
in Ref.[\refcite{higgsless}] is connected to gauge invariance. 
We should emphasize that the supercharge $Q$ is well defined for 
all the boundary conditions (\ref{bc_gauge_1})-(\ref{bc_gauge_4}) 
and hence that the supersymmetric structure always appears 
in the spectrum, though the boundary conditions other than 
the type (N,N) break 4d gauge symmetries, as we will see below.

From the above analysis, the 5d gauge fields $A_{\mu}(x,y)$ and 
$A_{y}(x,y)$ are expanded in the mass eigenstates as follows:
%
\begin{eqnarray}
	&& A_{\mu}(x,y) =\sum_n A_{\mu,n}(x)f_n(y) , \\
	&&A_{y} (x,y)=\sum_n h_n(x)g_n(y) ,
\end{eqnarray}
%
where $f_n(y)$ and $g_n(y)$ are the eigenstates of the 
Schr{\" o}dinger-like equations 
%
\begin{eqnarray}
\label{eigeneq_f_gauge}
	&&-\frac{1}{\sqrt{g_{55}}}\partial_y \frac{e^{-4W}}{\sqrt{g_{55}}}\partial_y f_n(y)=m_n^2 f_n(y) , \\
\label{eigeneq_g_gauge}
	&& -\partial_y \frac{1}{\sqrt{g_{55}}}\partial_y  \frac{e^{-4W}}{\sqrt{g_{55}}}g_n(y) = m_n^2 g_n(y) 
\end{eqnarray}
%
with one of the four types of the boundary conditions 
(\ref{bc_gauge_1})-(\ref{bc_gauge_4})
and they are actually related each other through the SUSY relations:
%
\begin{align}
\label{vectot_SUSY_relation_1}
m_{n} g_{n}(y) 
 &= \partial_{y}f_{n}(y), \\
\label{vectot_SUSY_relation_2}
m_{n} f_{n}(y)
 &= -\frac{1}{\sqrt{g_{55}}}\partial_{y}\frac{e^{-4W}}{\sqrt{g_{55}}}g_{n}.
\end{align}
%
Since the massless states are especially important in phenomenology, 
let us investigate the massless states of the equations 
(\ref{eigeneq_f_gauge}) and (\ref{eigeneq_g_gauge}).
Thanks to supersymmetry, the massless modes would be the solutions 
to the first order differential equation $Q | \Psi_0 \rangle = 0$, i.e.
%
\begin{eqnarray}
	&& \partial_y f_0(y) = 0, \\
	&&\partial_y \left( \frac{e^{-4W}}{\sqrt{g_{55}}}g_0(y) \right)=0.
\end{eqnarray}
%
The solutions are easily found to be 
%
\begin{eqnarray}
	&&f_0(y)=N_{0}, \\
	&& g_0(y) =\bar{N}_{0}\, e^{4W(y)}\sqrt{g_{55}(y)},
\end{eqnarray}
%
where $N_{0}$ and $\bar{N}_{0}$ are some constants.
We should emphasize that the above solutions do not necessarily 
imply physical massless states of $A_{\mu,0}(x)$ and $h_0(x)$ in the spectrum.
This is because the boundary conditions exclude some or all of 
them from the physical spectrum.
Indeed, $f_0(y)$ $(g_0(y))$ satisfies only the boundary conditions 
of the type (N,N) (type (D,D)).
Thus, a massless vector $A_{\mu,0}(x)$ (a massless scalar $h_0(x)$) 
appears only for the type (N,N) (type (D,D)) boundary conditions
(see Fig.4).
%
%
%
\begin{figure}
\begin{center}
\psfig{file=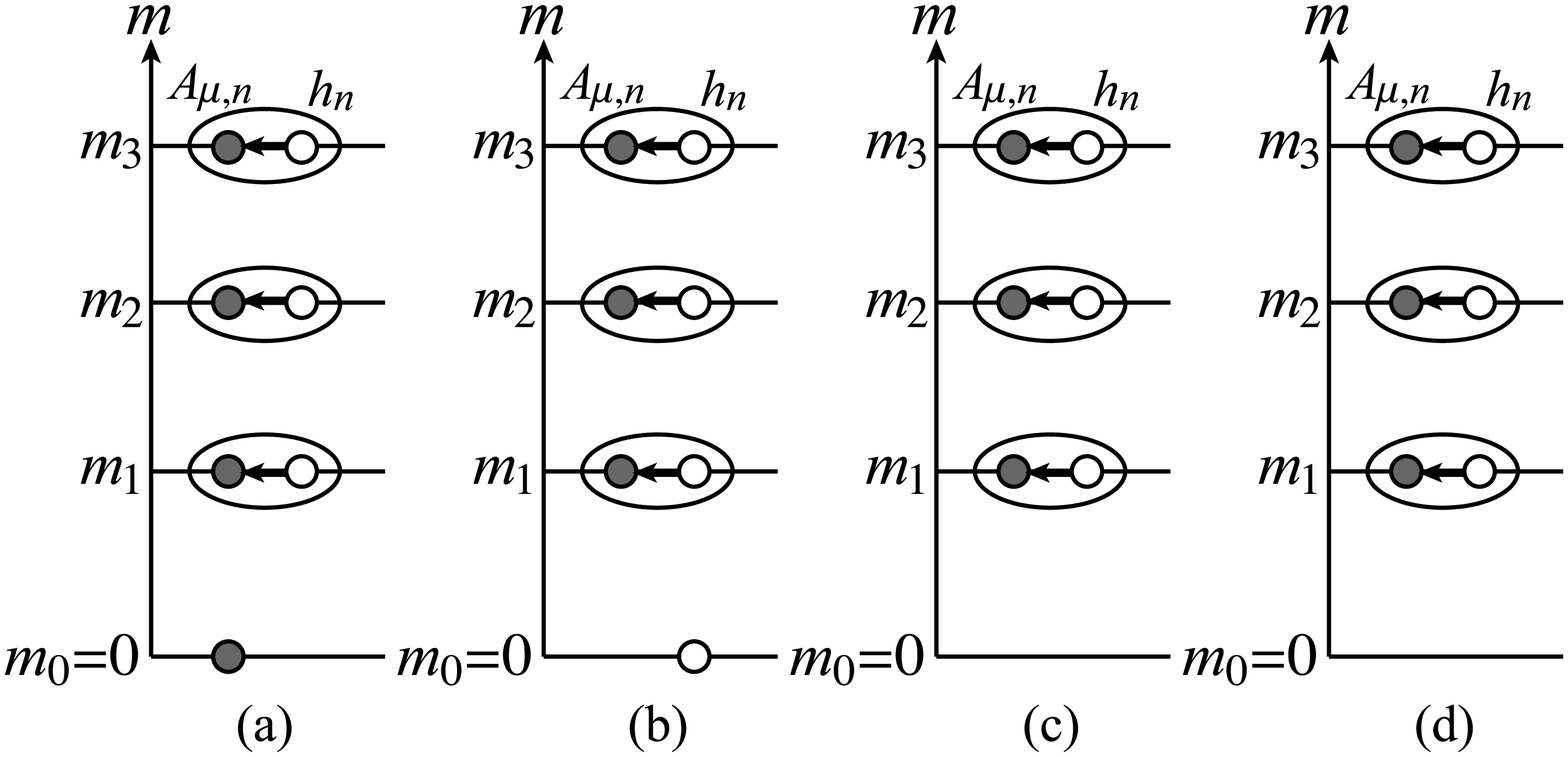,scale=0.35}
\end{center}
\vspace{-5mm}
\caption{Spectrum of the 5d gauge theory (a) for the type (N,N),
(b) for the type (D,D), (c) for the type (N,D),
(d) for the type (D,N) boundary conditions.
}
\label{fig4}
\end{figure}
%
%
%
%
%
%
This implies that the 4d gauge symmetry is broken except for 
the type (N,N) boundary conditions.

It is instructive to discuss the relation between the QM SUSY
and the higher dimensional gauge symmetry.
The relation becomes apparent by expressing the action, in terms
of the  4d mass eigenstates, as 
%
\begin{align}
\label{vectot_4d_action}
S = \int d^{4}x \bigl\{ {\cal L}_{m=0} + {\cal L}_{m\ne 0}\bigr\},
\end{align}
%
where ${\cal L}_{m=0}$ is the Lagrangian consisting of the massless
fields and
%
\begin{align}
\label{vectot_4d_massive_L}
{\cal L}_{m\ne 0}
 = \sum_{m_{n}\ne 0}\Bigl\{
   -\frac{1}{4}\bigl( F_{\mu\nu,n}(x) \bigr)^{2}
   -\frac{m_{n}^{\ 2}}{2}\Bigl( A_{\mu,n}(x)
    -\frac{1}{m_{n}}\partial_{\mu}h_{n}(x) \Bigr)^{2} \Bigr\}
\end{align}
%
with $F_{\mu\nu,n}=\partial_{\mu}A_{\nu,n}-\partial_{\nu}A_{\mu,n}$.
It follows that every nonzero mode $h_{n}(x)$ for $m_{n}\ne 0$
can be absorbed into the longitudinal mode of $A_{\mu,n}(x)$
and then $A_{\mu,n}(x)$ becomes massive with three degrees of
freedom, as it should be.
The choice of $h_{n}(x)=0\ (m_{n}\ne 0)$ is called a unitary
gauge.
It should be emphasized that the Lagrangian (\ref{vectot_4d_massive_L}) 
has been derived by use of the SUSY relations (\ref{vectot_SUSY_relation_1})
and (\ref{vectot_SUSY_relation_2}).
Therefore, the QM SUSY is necessary for $A_{\mu,n}(x)\ (m_{n}\ne 0)$ to
become massive by absorbing the unphysical mode $h_{n}(x)$,
 which is a consequence of the higher dimensional gauge symmetry.
This observation is summarized in Fig.4.

We have restricted our considerations to a 5d gauge theory.
The extension to any higher dimensional gauge theory is
possible and QM SUSY is found in the 4d mass spectrum.
The details have been given 
in Ref.[\refcite{SUSY_gauge_1}, \refcite{SUSY_gauge_2}].

%
%
%
\section{5d Gravity}
%
%
%
In this section, we investigate the 4d mass spectrum of the 5d
Randall-Sundrum gravity theory\cite{rs} with a warped metric
%
\begin{align}
\label{warped_metric}
ds^{2} = e^{2A(y)} \bigl(
         \eta_{\mu\nu} dx^{\mu}dx^{\nu} + dy^{2}\bigr),
\end{align}
%
where $A(y)$ is the warp factor which turns out to play
a role of a superpotential in the $N=2$ Witten model\cite{Witten_model}.
For the Randall-Sundrum model, the warp factor is given by
%
\begin{align}
\label{RS_warp_factor}
A(y) = - \ln \Bigl(\frac{y}{y_{1}}\Bigr).
\end{align}
%
Here, the location of the UV brane is chosen such that the
warp factor is set equal to $1$ on the UV brane at $y=y_{1}$.

The metric fluctuations $h_{MN}$ around the background metric
(\ref{warped_metric}) are given by
%
\begin{align}
\label{metric_fluctuation_1}
ds^{2}
 = e^{2A(y)} \bigl(
   \eta_{MN} + h_{MN}(x,y) \bigr) dx^{M}dx^{N}
\end{align}
%
and $h_{MN}$ turn out to be useful with the parameterization\cite{5d_gravity}
%
\begin{align}
\label{metric_fluctuation_2}
h_{MN}(x,y)
 = \left(
   \begin{array}{cc}
    h_{\mu\nu}(x,y) - \frac{1}{2}\eta_{\mu\nu}\phi(x,y)
    & \ \ h_{y\nu}(x,y)\\
    h_{\mu y}(x,y) & \ \ \phi(x,y)
   \end{array} \right).
\end{align}
%
The action is invariant under infinitesimal general coordinate
transformations: $x^{M} \to x^{\prime M} = x^{M} + \xi^{M}(x,y)$,
which are translated into the field transformations of the
metric fluctuations:
%
\begin{align}
\label{general_coordinate_transformation}
\delta h_{\mu\nu}
 &= -\partial_{\mu}\xi_{\nu} - \partial_{\nu}\xi_{\mu}
    - \eta_{\mu\nu} (\partial_{y}+3A^{\prime})\xi_{y},\notag\\
\delta h_{\mu y}
 &= -\partial_{y}\xi_{\mu} - \partial_{\mu}\xi_{y},\notag\\
\delta\phi
 &= -2(\partial_{y} + A^{\prime}) \xi_{y},
\end{align}
%
where $A'(y)=dA(y)/dy$.
The metric fluctuation fields are expanded, in terms of some
complete sets of the functions $\{f_{n}(y)\},\ \{g_{n}(y)\},\ 
\{k_{n}(y)\}$, as 
%
\begin{align}
\label{mode_expansion_metric_fluctuation}
h_{\mu\nu}(x,y)
 &= \sum_{n}h_{\mu\nu, n}(x) f_{n}(y),\notag\\
h_{\mu y}(x,y)
 &= \sum_{n}h_{\mu y, n}(x) g_{n}(y),\notag\\
\phi(x,y)
 &= \sum_{n}\phi_{n}(x) k_{n}(y). 
\end{align}
%
It is instructive to examine degrees of freedom for massive modes.
Each field of $h_{\mu\nu,n}, h_{\mu y,n}, \phi$ has originally
$2, 2, 1$ degrees of freedom, respectively because 5d gravity has
no mass term in a 5-dimensional point of view.
The vector field $h_{\mu y, n}$ could become massive by \lq\lq eating"
one extra degree of freedom and then has three degrees of freedom as
a massive vector.
The tensor field $h_{\mu\nu,n}$ could become massive by \lq\lq eating"
three extra degrees of freedom and then has five degrees of freedom
as a massive graviton.
The above Higgs-like mechanism can actually occur in the 5d gravity
system.
The vector field $h_{\mu y,n}$ \lq\lq eats" $\phi_{n}$ to become
a massive vector with three degrees of freedom, and then the tensor
field $h_{\mu\nu,n}$ \lq\lq eats" $h_{\mu y,n}$ to become a massive
graviton with five degrees of freedom:
%
\begin{align}
\label{geometrical_Higgs_mechanism}
\overbrace{h_{\mu\nu,n}\quad
 \underbrace{h_{\mu y,n}\quad \ 
 \phi_{n}}_{\textrm{a massive vector}} }^{\textrm{a massive graviton}}
\end{align}
%

The above observation strongly suggests, on the analogy of the
5d gauge theory, that the 5d gravity theory possesses {\em two}
QM SUSY systems in the 4d spectrum:
One is realized between the eigenfunctions $g_{n}(y)$ and $k_{n}(y)$.
The other is between $f_{n}(y)$ and $g_{n}(y)$.
To verify it, we have to find the eigenequations for
$f_{n}(y), g_{n}(y)$ and $k_{n}(y)$, which should diagonalize 
the quadratic action for $h_{\mu\nu,n}(x), h_{\mu y,n}(x)$ and
$\phi_{n}(x)$.
The eigenequations for $f_{n}(y), g_{n}(y)$ 
and $k_{n}(y)$ are found to be\cite{5d_gravity}
%
\begin{align}
\label{eigenequations_f}
-\bigl(\partial_{y}^{\ 2} + 3A'(y)\partial_{y}\bigr) f_{n}(y)
 &= m_{n}^{\ 2}f_{n}(y),\\
\label{eigenequations_g}
-\bigl(\partial_{y}^{\ 2} + 3A'(y)\partial_{y}
 + 3A''(y)\bigr) g_{n}(y)
 &= m_{n}^{\ 2}g_{n}(y),\\
\label{eigenequations_k}
-\bigl(\partial_{y}^{\ 2} + 3A'(y)\partial_{y}
 + 4A''(y)\bigr) k_{n}(y)
 &= m_{n}^{\ 2}k_{n}(y).
\end{align}
%
The supersymmetric structure between $f_{n}(y)$ and $g_{n}(y)$
will become apparent if we express Eqs.(\ref{eigenequations_f})
and (\ref{eigenequations_g}) into the form
%
\begin{align}
\label{eigenequations_f&g}
{\cal D}^{\dagger} {\cal D} f_{n}(y) = m_{n}^{\ 2}f_{n}(y),\quad
{\cal D} {\cal D}^{\dagger} g_{n}(y) = m_{n}^{\ 2}g_{n}(y),
\end{align}
%
where 
%
\begin{align}
\label{def_D&D+}
{\cal D} = \partial_{y},\quad 
{\cal D}^{\dagger} = -\bigl(\partial_{y} + 3A'(y)\bigr).
\end{align}
%
The eigenfunctions $f_{n}(y)$ and $g_{n}(y)$ are actually
related each other through the SUSY relations
%
\begin{align}
\label{SUSY_relation_f&g}
m_{n}g_{n}(y) = {\cal D}f_{n}(y), \quad
m_{n}f_{n}(y) = {\cal D}^{\dagger} g_{n}(y).
\end{align}
%
It seems strange that ${\cal D}^{\dagger}$ is hermitian conjugate
to ${\cal D}$.
This is, however, true because the inner product 
is defined by\cite{5d_gravity}
%
\begin{align}
\label{def_inner_product}
\langle\, \psi \vert \phi\,\rangle
 = \int_{y_{2}}^{y_{2}}dy e^{3A(y)} \bigl(\psi(y)\bigr)^{*} \phi(y)
\end{align}
%
with the boundary conditions
%
\begin{align}
\label{bc_f&g}
\partial_{y} f_{n}(y) = 0 = g_{n}(y),\quad \textrm{at}\ 
y=y_{1}, y_{2}.
\end{align}
%
The factor $e^{3A(y)}$ in Eq.(\ref{def_inner_product}) is
required because of the presence of it in the action,
whose origin comes from the nontrivial background metric
(\ref{warped_metric}).
The boundary conditions (\ref{bc_f&g}) turn out to be 
compatible with supersymmetry.
The QM SUSY structure will be manifest if we introduce two
component wavefunctions
%
\begin{align}
\label{2-component_wavefunction_gravity}
\vert \Psi \rangle 
 = \left(
    \begin{array}{c}
     f(y)\\
     g(y)
    \end{array} \right).
\end{align}
%
Then, $H, Q$ and $(-1)^{F}$ are found to be the same
form as Eq.(\ref{def_H_spinor}) with 
${\cal D}$ and ${\cal D}^{\dagger}$ defined in Eqs.(\ref{def_D&D+}).

Let us next proceed to the analysis of a pair of the
eigenfunctions $g_{n}(y)$ and $k_{n}(y)$.
The supersymmetric structure between them will be apparent if
we express Eqs.(\ref{eigenequations_g}), (\ref{eigenequations_k})
into the form\cite{5d_gravity}
%
\begin{align}
\label{eigenequations_g&k}
\bar{{\cal D}}^{\dagger} \bar{{\cal D}} g_{n}(y) 
 = m_{n}^{\ 2}g_{n}(y),\quad
\bar{{\cal D}} \bar{{\cal D}}^{\dagger} k_{n}(y) = m_{n}^{\ 2}k_{n}(y),
\end{align}
%
where
%
\begin{align}
\label{def_barD&D+}
\bar{{\cal D}} = \partial_{y} + A'(y),\quad 
\bar{{\cal D}}^{\dagger} = -\bigl(\partial_{y} + 2A'(y)\bigr).
\end{align}
%
Here, we have used the relation $(A')^{2}=A''$.
The eigenfunctions $g_{n}(y)$ and $k_{n}(y)$ are related
through the SUSY relations
%
\begin{align}
\label{SUSY_relation_g&k}
m_{n}k_{n}(y) = \bar{{\cal D}}g_{n}(y), \quad
m_{n}g_{n}(y) = \bar{{\cal D}}^{\dagger} k_{n}(y).
\end{align}
%
The inner product is defined by Eq.(\ref{def_inner_product}).
This guarantees that $\bar{{\cal D}}$ and $\bar{{\cal D}}^{\dagger}$
are hermitian conjugate each other with the boundary conditions
%
\begin{align}
\label{bc_g&k}
g_{n}(y) = 0 = \bar{{\cal D}}^{\dagger}k_{n}(y),\quad \textrm{at}\ 
y=y_{1}, y_{2}.
\end{align}
%
The supersymmetric structure is manifest if we introduce two component
wavefunctions in a similar manner as Eq.(\ref{2-component_wavefunction_gravity}).
Then, $\bar{H}, \bar{Q}$ and $(-1)^{\bar{F}}$ are given by the same
form as Eq.(\ref{def_H_spinor}) with the replacement of 
${\cal D}$ and ${\cal D}^{\dagger}$ by Eqs.(\ref{def_barD&D+}).

We have found two QM SUSY systems, as expected.
The eigenfunctions $f_{n}(y)$ and $g_{n}(y)$ form a supermultiplet,
and $g_{n}(y)$ and $k_{n}(y)$ form another supermultiplet.
The QM SUSY structure turns out to severely restrict the allowed
boundary conditions for $f_{n}(y), g_{n}(y)$ and $k_{n}(y)$.
In fact, the boundary conditions for them are unique in order to 
be compatible with the QM SUSY.
This fact is especially important in a low energy effective theory
point of view.
This is because the boundary conditions for $f_{n}(y), g_{n}(y)$ 
and $k_{n}(y)$ with the eigenvalue equations (\ref{eigenequations_f&g}) 
and (\ref{eigenequations_g&k}) determine uniquely the massless modes.
In the present Randall-Sundrum model, there exist one massless
graviton  and one massless scalar (radion).
The 4d spectrum of the Randall-Sundrum model is depicted
in Fig.5.
%
%
%
\begin{figure}
\begin{center}
\psfig{file=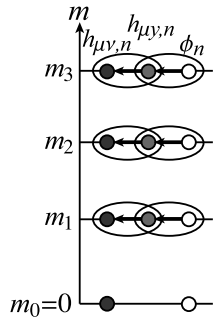,scale=0.35}
\end{center}
\vspace{-5mm}
\caption{Spectrum of the 5d Randall-Sundrum gravity}
\label{fig5}
\end{figure}
%
%
%

We have investigated the 5d Randall-Sundrum model in which the
4-dimensional space-time is a flat Minkowski.
Karch and Randall\cite{KR} have extended it to 4d de Sitter ($dS_{4}$)
and anti de Sitter ($AdS_{4}$) space-time.
In those cases, the warp factor $A(y)$ is different from Eq.
(\ref{RS_warp_factor}) with a non-vanishing 4d cosmological
constant.
The analysis proceeds in a similar way and the results will
be reported elsewhere.

Finally, we should make a few comments on interesting observations.
The warp factor $A(y)$ cannot be an arbitrary function but has to be
a solution of the Einstein equation.
This gives a non-trivial constraints on $A(y)$.
The differential equations for $f_{n}(y), g_{n}(y)$ and $k_{n}(y)$
are found to be in a class of exactly solvable models with
the property of shape invariance\cite{exactly_solvable}.
This property holds even for the Karch-Randall models.
The second interesting observation is the uniqueness of the
boundary conditions which have to be compatible with two
QM SUSYs.
If we would have a 5d massless theory with a higher spin
$(s>2\hbar)$, the 4d mass spectrum of the system could possess
more than three QM SUSYs.
Our analysis, however, tells us that there are no possible 
boundary conditions compatible with all QM SUSYs.
This may lead to a conclusion that any 5d massless theory on an
interval with higher spins $(s>2\hbar)$ has no possible boundary
conditions compatible with QM SUSYs.
This seems to be consistent with the fact that any non-trivial
massless higher spin theory with $s>2\hbar$ has not been found yet.

%
%
%
\section{Conclusions}
%
%
%
We have investigated higher dimensional scalar, spinor, gauge and
gravity theories from a 4d spectrum point of view.
Our analysis has shown that QM SUSY is hidden in 4d mass spectrum
of any higher dimensional field theories except for scalars.
The origins of QM SUSYs in the 4d mass spectrum are found to
be chiral symmetry, higher dimensional gauge symmetry and higher 
dimensional general covariance symmetry for spinor, gauge and
gravity theories, respectively.
There is no such symmetry to guarantee QM SUSY in the 4d
mass spectrum for scalar theories.
QM SUSY could appear in higher dimensional scalar theories but
it is an accidental symmetry.

The higher dimensional gauge invariance guarantees that
the nonzero vector mode $A_{\mu,n}(x)\ (n>0)$ can absorb the
unphysical scalar mode $h_{n}(x)$ to become massive with three
degrees of freedom.
This is the origin of QM SUSY between the mass eigenfunctions
$f_{n}(y)$ and $g_{n}(y)$ for $A_{\mu,n}(x)$ and $h_{n}(x)$.

The higher dimensional general covariance symmetry similarly
guarantees that the nonzero vector mode 
$h_{\mu y,n}(x)\ (n>0)$ can absorb the unphysical scalar mode
$\phi_{n}(x)$ to become massive with three degrees of freedom,
and that the nonzero graviton mode $h_{\mu\nu,n}(x)\ (n>0)$
can then absorb the massive mode $h_{\mu y,n}(x)$ to become
massive with five degrees of freedom.
This is the origin of QM SUSY and there appear {\em two} QM SUSYs
in higher dimensional gravity theories:
One connects the mass eigenfunction $g_{n}(y)$ to $k_{n}(y)$.
The other connects the mass eigenfunction $f_{n}(y)$ to $g_{n}(y)$.

It is interesting to point out that the chiral, higher dimensional
gauge and higher dimensional general covariance symmetries
are all related to the symmetries that guarantee the masslessness
of spinor, vector and tensor fields, respectively.
Since massless particles are crucially important at low energy
physics, it would be of great interest to investigate QM SUSY
in more details.
Our results will be summarized in the Table 1.
%
%
\begin{table}
\tbl{Summary of our results}
{\begin{tabular}{@{}ccc@{}}\toprule
higher dim. fields & QM SUSY & origin \\\colrule
scalar & $\bigtriangleup$ & accidental \\
spinor & $\bigcirc$ & chiral symmetry \\
vector & $\bigcirc$ & higher dim. gauge symmetry \\
tensor & $\bigcirc$ & higher dim. general covariance symmetry \\\botrule
\end{tabular}}
\label{aba:tbl1}
\end{table}
%
%

\section*{Acknowledgments}
This work has been supported in part by a Grant-in-Aid 
for Scientific Research (No. 22540281) from the Japanese 
Ministry of Education, Science, Sports and Culture. 
The author would like to thank Y. Fujimoto, C.S. Lim,
T. Nagasawa, S. Ohya, K. Sakamoto, K. Sekiya, H. Sonoda, 
K. Takenaga for valuable discussions.


%
%
%

%
%
%
\end{document}